\documentstyle[11pt,psfig,graphics]{article}

\newcommand{\ds}{\displaystyle }
\newcommand{\R}{{\sf R\hspace*{-0.9ex}%
\rule{0.15ex}{1.5ex}\hspace*{0.9ex}}}
\newcommand{\Z}{{\sf Z\hspace*{-0.9ex}%
\rule{0.15ex}{1.5ex}\hspace*{0.9ex}}}

\title{Relations de dispersion pour cha\^{i}nes lin\'{e}aires comportant des interactions harmoniques auto-similaires}

\author{{\sl\small T.M. Michelitsch $^{1}$\footnote{Corresponding author, Email:
michel@lmm.jussieu.fr } \, G.A. Maugin$^{1}$ F. C. G. A. Nicolleau$^2$, A. F. Nowakowski$^2$, S. Derogar$^3$} \\[5pt]
 $^1$ Institut Jean le Rond d'Alembert \\ CNRS UMR 7190 \\
 Universit\'{e} Pierre et Marie Curie, Paris 6 \\ France
 \\[5pt]
 \\
$^2$ Department of Mechanical Engineering, University of Sheffield, Royaume-Uni \\
$^3$ Department of Civil and Structural Engineering,
University of Sheffield, Royaume-Uni
\\[5pt] \\ {\it filename: \jobname .tex }
\\[5pt]
 \\
{\bf d\'{e}pos\'{e} sur HAL} }

\begin{document}

\maketitle


\section{R\'{e}sum\'{e}}

{\small Dans de nombreux syst\`{e}mes biologiques on trouve des structures arborescentes et bifurquantes
comme les poumons, les arbres, les foug\`{e}res, les coquilles des escargots, le syst\`{e}me vasculaire, etc. Tous ces syst\`{e}mes se distinguent par une invariance d'\'{e}chelle et se comportent donc quasiment comme des syst\`{e}mes auto-similaires. Est-ce par soucis d'optimisation qu'une telle sym\'{e}trie a \'et\'e retenue par la nature\,? Quels sont les comportements dynamiques et acoustiques de tels syst\`{e}mes auto-similaires\,? Afin de donner une r\'{e}ponse il faut tout d'abord comprendre le r\^ole de l'auto-similarit\'{e} dans les comportements dynamiques comme la propagation d'ondes. Faute d'approches d\'{e}finitives qui d\'{e}crivent la physique de syst\`{e}mes auto-similaires, on introduit ici le mod\`{e}le simple d'une cha\^{i}ne lin\'{e}aire quasi-continue correspondant \`a une distribution auto-similaire de ressorts qui lient des masses ponctuelles entre-elles. Dans notre mod\`{e}le la densit\'{e} d'\'{e}nergie \'{e}lastique est une fonction exactement auto-similaire. L'\'{e}quation du mouvement fournit une version auto-similaire de l'op\'{e}rateur de Laplace dont le spectre de valeurs propres (la relation de dispersion) est une fonction de type Weierstrass-Mandelbrot. Cette fonction est exactement auto-similaire et est de plus une fonction fractale non-diff\'{e}rentiable.

En outre, il s'av\`{e}re que dans le cas limite d'un milieu continu notre Laplacien adopte la forme d'int\'{e}grales fractionnelles. Dans ce cas la densit\'e de vibrations se r\'{e}v\`{e}le \^etre une loi de puissance pour les basses fr\'{e}quences avec un exposant strictement positif qui indique l'annulation de la densit\'{e} de vibrations pour une fr\'{e}quence nulle.
Pour de plus amples d\'etails nous renvoyons
le lecteur \`a notre article r\'ecent \cite{michel-et-al}.
\newline\newline\newline
\noindent {\bf Mots-cl\'{e}s:} Auto-similarit\'{e}, fonctions auto-similaires, transformations affines,
fonction de Weierstrass-Mandelbrot, fonctions fractales, loi de puissance, int\'{e}grales fractionnelles.}
\newline\newline
\noindent {\bf PACS: 05.50.+q 81.05.Zx 63.20.D-}

\section{Introduction}
\label{intro}

Ce fut dans les ann\'ees soixante-dix du dernier si\`ecle que Beno\^it Mandelbrot
d\'eclencha une r\'evolution scientifique en d\'eveloppant sa {\it G\'eom\'etrie Fractale} \cite{1}.
Toutefois, les orgigines de cette discipline remontent au moins jusqu'au 19{\`eme} si\`ecle \cite{8}.
Depuis lors, les applications des structures fractales se d\'eveloppent de plus en plus, citons par exemple les antennes fractales \cite{28,24}. En outre, des r\'eseaux
de type Sierpinski r\'ev\`elent des caract\'eristiques vibratoires int\'eresantes \cite{3}.

Tandis que des syst\`emes fractals et auto-similiares sont d\'ej\`a un sujet d'\'etudes bien \'etabli dans de nombreux domaines de la physique, ils attirent aussi de plus en plus d'inter\^et en m\'ecanique analytique et dans les sciences de l'ing\'enieur.
Il faut se rendre compte qu'il y a encore une large m\'econnaissance
de l'impact de l'auto-similarit\'e en tant que sym\'etrie sur les comportements physiques.
Cependant quelques pas initiaux ont d'ores et d\'ej\`a \'et\'e effectu\'es \cite{3,2,tara,6,7,epstein},
alors que le developpement d'une {\it m\'ecanique fractale} se fait attendre.
Faute d'une th\'eorie d\'efinitive, il est donc d\'esirable d'\'etablir des mod\`eles de syt\`emes auto-similaires qui soient \`a la fois pertinents et autant que possible simples \`a analyser. C'est donc l'objectif de cet article que de d\'evelopper un tel mod\`ele.

Nous explorons les comportements vibratoires d'une cha\^ine lin\'eaire ayant
des interactions inter-particulaires harmoniques r\'ealisant une distribution spatiale auto-similaire.
Quelques \'etudes ont \'et\'e consacr\'ees aux cha\^ines lin\'eaires avec aussi bien des propri\'et\'es fractales
\cite{tara,prl9122,prl3297,prl1239,prl3110} que des propri\'et\'es en d\'ependance exponentielle \cite{michelmaugin}.

Au contraire de ces articles consacr\'es aux grilles discr\`etes, nous analysons ici les comportements oscillatoires
dans une cha\^ine lin\'eaire qui est non seulement fractale mais aussi exactement {\it auto-similaire} par rapport \`a ses interactions inter-particulaires. Il semble que notre approche pluridisciplinaire soit utile
pour la mod\'elisation de certains probl\`emes en m\'ecanique des fluides, notamment en milieux turbulents \cite{weiermandel,franck}.

La d\'emonstration est d\'evelopp\'ee comme suit~: le \S~\ref{thomas1} est consacr\'e \`a la construction
de fonctions et op\'erateurs auto-similaires. Avec cette machinerie on d\'eduit une sorte de Laplacien auto-similaire.
Bien que la construction d'un tel op\'erateur ne soit pas unique, on d\'emontre pourtant
que cette mani\`ere de construction dispose d'une certaine justification physique. Dans une approximation
continue notre Laplacien rev\^et la forme d'int\'egrales fractionnelles.

Dans le \S~\ref{thomas2} on \'etablit un mod\`ele physique simple de cha\^ine lin\'eaire de longueur infinie comportant des interactions harmoniques (ressorts) auto-similaires. L'auto-similarit\'e requiert une distribution continue et homog\`ene des particules. L'\'equation de mouvement appara\^it sous forme d'une \'equation d'ondes
auto-similaire qui contient notre Laplacien auto-similaire. Les valeurs propres de ce Laplacien
fournissent la r\'elation de dispersion qui est une fonction de type {\it Weierstrass-Mandelbrot} qui comporte
aussi bien l'auto-similarit\'e que des propri\'et\'es fractales.

\section{Definition du probl\`eme affin}
\label{thomas1}

Tout d'abord nous d\'efinissons la notion d'``auto-similarit\'e" en termes de fonctions et op\'erateurs.
Nous d\'enommons une fonction scalaire $\phi(h)$ {\it
exactemant auto-similaire} par rapport \`a la variable $h$ si la condition
\begin{equation}
\label{self-sim}
\phi(Nh)=\Lambda\phi(h)
\end{equation}
est satisfaite pour toutes valeurs $h>0$. Nous d\'enommons (\ref{self-sim}) le ``probl\`eme affin"\footnote{nous nous restreignons aux transformations affines de la forme $h'=Nh+c$ with $c=0$.}. $N$ est un param\`etre fixe pour un probl\`eme donn\'e et $\Lambda=N^{\delta}$ un ensemble continu de valeurs propres dont la bande admissible de l'exposant $\delta=\frac{\ln{\Lambda}}{\ln{N}}$ est \`a d\'eterminer.
Une fonction qui satisfait (\ref{self-sim}) pour un certain $N$ et une bande $\Lambda=N^{\delta}$
repr\'esente une solution $\phi_{N,\delta}(h)$ du probl\`eme (\ref{self-sim}) dont
le param\`etre fixe $N$ est donn\'e. Il va de soi que si une fonction $\phi(h)$ satisfait
(\ref{self-sim}) elle satisfait aussi $\phi(N^sh)=\Lambda^s\phi(h)$ o\`u $s \in \Z$ (z\'ero inclus).
En choisissant $s=-1$, il est clair que nous r\'ecuperons le m\^eme probl\`eme en
remplacant $N$ et $\Lambda$ par $N^{-1}$ et $\Lambda^{-1}$ dans (\ref{self-sim}).
Il s'ensuit que nous pouvons restreindre notre analyse aux valeurs $N>1$ sans aucune perte de
g\'en\'eralit\'e. De plus, il est justifi\'e pour des raisons physiques d'admettre $\Lambda, N$
comme nombres r\'eels et positifs ($\Lambda, N \in \R^{+}$).

Pour aborder le probl\`eme affin (\ref{self-sim}) comme probl\`eme de valeurs propres, il convient
d'introduire l'op\'erateur affin ${\hat A}_{N}$ comme suit

\begin{equation}
\label{affinetrafo} {\hat A}_{N}f(h)=: f(Nh)
\end{equation}

On v\'erifie facilement que l'op\'erateur affin ${\hat A}_{N}$ est lin\'eaire, c'est \`a dire que
\begin{equation}
\label{linearite} {\hat
A}_{N}\left(c_1f_1(h)+c_2f_2(h)\right)=c_1f_1(Nh)+c_2f_2(Nh)
\end{equation}
et encore
\begin{equation}
\label{pouvoir} {\hat A}_{N}^sf(h)=f(N^sh),\hspace{2cm}
s=0\pm1,\pm2, ...\pm\infty
\end{equation}

Avec cela on peut d\'efinir des fonctions d'op\'erateurs. Soit $g(\tau)$ une fonction scalaire
assez lisse pour \^etre d\'evelopp\'ee en s\'erie de McLaurin comme suit
\begin{equation}
\label{Taylor} g(\tau)=\sum_{s=0}^{\infty}a_s\tau^s
\end{equation}
nous d\'efinissons une fonction de l'op\'erateur affin correspondant \`a $g$ par
la s\'erie
\begin{equation}
\label{fonctionop} g(\xi{\hat A}_N)=\sum_{s=0}^{\infty}a_s\xi^s{\hat
A}_N^s
\end{equation}
o\`u $\xi$ repr\'esente un param\`etre scalaire. Cette fonction de l'oper\'erateur affin
agit sur une fonction $f(h)$ scalaire comme suit
\begin{equation}
\label{fonctionopact} g(\xi{\hat
A}_N)f(h)=\sum_{s=0}^{\infty}a_s\xi^sf(N^sh)
\end{equation}

La convergence de cette s\'erie est \`a v\'erifier pour toute fonction $f(h)$ donn\'ee. Il est aussi utile d'exprimer l'op\'erateur affin en utilisation la notation explicite
\begin{equation}
\label{affineexplicit} {\hat A}_N(h)=e^{\ln N\frac{\rm d}{\rm d (\ln
h)}}
\end{equation}

Par ce formalisme nous sommes \`a m\^eme de construire les fonctions et op\'erateurs
auto-similaires n\'ecessaires \`a l'analyse, dans la \S~\ref{thomas2}, du probl\`eme physique
d'une cha\^ine lin\'eaire auto-similaire.

\subsection{Construction de fonctions et op\'erateurs auto-similaires}

Une solution du probl\`eme affin (\ref{self-sim}) s'\'ecrit

\begin{equation}
\label{self-sim-func}
\phi(h)=\sum_{s=-\infty}^{\infty}\Lambda^{-s}{\hat A}_{N}^sf(h)=\sum_{s=-\infty}^{\infty}\Lambda^{-s}f(N^sh)
\end{equation}
pour toute fonction $f(h)$ dont la s\'erie (\ref{self-sim-func})
converge r\'eguli\`erement pour tous $h$.
On introduit l'op\'erateur auto-similaire comme suit

\begin{equation}
\label{selfsimop}
{\hat
T}_N=\sum_{s=-\infty}^{\infty}\Lambda^{-s}{\hat A}_{N}^s
\end{equation}
qui satisfait la condition d'auto-similarit\'e ${\hat
A}_N{\hat T}_N=\Lambda {\hat T}_N$. Nous soulignons le fait que
nous pouvons nous restreindre aux cas $N>1$ ($N,\Lambda \in \R$)
et exclure le cas pathologique $N=1$.

Il est important de pr\'eciser pour quelles fonctions $f(t)$ la s\'erie (\ref{self-sim-func})
est convergente. On trouve que pour qu'une fonction $f(t)$ soit admissible, elle doit satisfaire les conditions ($t>0$)\cite{michel-et-al}
\begin{equation}
\label{ass1} \lim_{t\rightarrow 0} f(t)=a_0\,t^{\alpha}
\end{equation}
et
\begin{equation}
\label{ass2} \lim_{t\rightarrow \infty} f(t)=c_{\infty}\, t^{\beta}
\end{equation}
o\`u $a_0, c_{\infty}$ sont des constantes. Les deux exposants $\alpha,\beta \in \R$
peuvent rev\^etir des valeurs r\'eelles, positives ou n\'egatives avec $\beta < \alpha$.
Pour une telle fonction, la fonction $\phi(h)$ de (\ref{self-sim-func}) est convergente
et donc existe si l'exposant $\delta=\frac{\ln{\Lambda}}{\ln{N}}$ est constraint
par
\begin{equation}
\label{lam2}
\beta < \delta=\frac{\ln{\Lambda}}{\ln{N}} < \alpha
\end{equation}

Le cas $\beta=0$ comprend des fonctions born\'ees $|f(t)| < M$ ci-inclus
quelques fonctions p\'eriodiques.

\subsection{Une variante auto-similaire de l'op\'erateur de Laplace}

Dans l'esprit de (\ref{self-sim-func}) et (\ref{selfsimop}) on construit
une fonction exactement auto-similaire telle que
\begin{equation}
\label{uhdef}
\phi(x,h)=
{\hat T}_N(h)\left(u(x+h)+u(x-h)-2u(x)\right)
\end{equation}
o\`u $u(..)$ signifie un champ continu arbitraire et suffissamment lisse,
et la d\'ependence de ${\hat T}_N(h)$ en variable $h$ exprime que
l'op\'erateur affin ${\hat A}_N(h)$ n'agit que sur $h$, de mani\`ere que ${\hat A}_N(h)v(x,h)=v(x,Nh)$. En posant $\xi=\Lambda^{-1}$
on a l'expression
\begin{equation}
\label{uh}
\phi(x,h)=
\sum_{s=-\infty}^{\infty}\xi^s\left\{u(x+N^sh)+u(x-N^sh)-2u(x)\right\}
\end{equation}
laquelle est auto-similaire par rapport \`a la variable $h$, soit
${\hat A}_N(h)\phi(x,h)$ $= \phi(x,Nh)=\xi^{-1}\phi(x,h)$. Par contre
elle est une fonction r\'eguli\`ere par rapport \`a la variable $x$.
De plus, on exige que le champ $u(x)$ soit Fourier transformable, ce qui impose comme condition
que l'int\'egrale
\begin{equation}
\label{uint} \int_{-\infty}^{\infty}|u(x)|\,{\rm d}x < \infty
\end{equation}
existe. Il est utile de noter que
\begin{equation}
\label{diffeq} u(x+h)+u(x-h)-2u(x)=4\sinh^2\left({\frac{h}{2}\frac{d}{dx}}\right)u(x)=h^2\frac{d^2}{dx^2}u(x)+ {\mathcal{O}} (\,\,{h^{4}})
\end{equation}
en utilisant $u(x\pm h)=e^{\pm h\frac{d}{dx}}u(x)$ avec
$u(x+h)+u(x-h)-2u(x)=\left(
e^{h\frac{d}{dx}}+e^{-h\frac{d}{dx}}-2\right)u(x)$. Ce qui est possible
puisque $u(x)$ est suppos\'ee suffissamment lisse.
De (\ref{uint}) et (\ref{diffeq}) il suit que la s\'erie (\ref{uh}) converge si l'exposant
$\delta$ se retrouve dans l'intervalle
\begin{equation}
\label{expon}
0 <  \delta=- \frac{\ln{\xi}}{\ln{N}} < 2
\end{equation}

Alors on introduit une variante du Laplacien auto-similaire de sorte que (avec $\xi=N^{-\delta}$)
\cite{michel-et-al}
\begin{equation}
\label{Laplaself3}
 \Delta_{(\delta,N,h)}u(x)=: \lim_{n\rightarrow \infty}N^{\delta
 n}\phi(x,N^{-n}h) =\phi(x,h)
 \end{equation}
o\`u $\phi(x,h)$ est la s\'erie (\ref{uh}). En utilisant (\ref{diffeq}),
le Laplacien (\ref{Laplaself3}) se r\'e\'ecrit comme une s\'erie d'op\'erateurs
\begin{equation}
\label{lapla}
 \Delta_{(\delta,N,h)}=4{\hat
 T}_N(h)\sinh^2\left({\frac{h}{2}\frac{\partial}{\partial
 x}}\right)=4\sum_{s=-\infty}^{\infty}N^{-\delta s}\sinh^2\left({\frac{N^sh}{2}\frac{\partial}{\partial
 x}}\right)
\end{equation}

Notamment on observe que la condition d'auto-similarit\'e
\begin{equation}
\label{selflap}
\Delta_{(\delta,N,Nh)}=N^{\delta}\Delta_{(\delta,N,h)}
\end{equation}
est en fait r\'espect\'ee par notre Laplacien.

\subsection{Approximation continue - lien avec les int\'egrales fractionnelles}
\label{contapprox}

Pour une \'evaluation num\'erique, il conviendrait d'employer une approximation
continue pour le Laplacien (\ref{lapla}). Pour cela on pose
$N=1+\epsilon$ (avec $0<\epsilon \ll 1$ de sorte que
$\epsilon\approx\ln{N}$) o\`u $\epsilon$ est un param\`etre suppos\'e petit.
On d\'efinit de plus $s\epsilon=v$ avec ${\rm d}v\approx \epsilon$ et
$N^s=(1+\epsilon)^{\frac{v}{\epsilon}}\approx e^v$.
Dans cette approche $N^s\approx e^v$ devient une variable quasiment continue
dont $s$ parcourt tous les $s\in \Z$.
(\ref{self-sim-func}) s'\'ecrit donc

\begin{equation}
\label{Laplacesum}\phi(h)=\sum_{s=-\infty}^{\infty}N^{-s\delta}f(N^sh)\approx
\frac{1}{\epsilon}\int_{-\infty}^{\infty}e^{-\delta v}f(h e^v){\rm d}v
\end{equation}
en posant $he^{v}=\tau$ ($h>0$) et
$\frac{{\rm d}\tau}{\tau}={\rm d}v$ avec $\tau(v\rightarrow
-\infty)=0$ et $\tau(v\rightarrow \infty)=\infty$, on trouve que

\begin{equation}
\label{fract} \phi(h)\approx
\frac{h^{\delta}}{\epsilon}\int_0^{\infty}\frac{f(\tau)}{\tau^{1+\delta}}\,{\rm
d}\tau
\end{equation}

L'application de cette relation au Laplacien (\ref{lapla}) donne
\begin{equation}
\label{fractlapself} \Delta_{(\delta,\epsilon,h)}u(x)\approx
\frac{h^{\delta}}{\epsilon}\int_0^{\infty}\frac{(u(x-\tau)+u(x+\tau)-2u(x))}{\tau^{1+\delta}}\,{\rm
d}\tau
\end{equation}

Cette int\'egrale existe pour $\beta < 0 < \delta < 2$ avec $\beta < -1$ du fait des conditions d'existence
de l'int\'egrale (\ref{uint}) et de la relation (\ref{diffeq}).
En effectuant deux int\'egrations partielles entre les limites $\tau=0$ et $\tau=\infty$
on r\'e\'ecrit (\ref{fractlapself}), \`a condition que $0 < \delta < 2$, sous la forme d'une convolution

\begin{equation}
\label{reforme} \Delta_{(\delta,\epsilon,h)}u(x)\approx
\int_{-\infty}^{\infty}g(|x-\tau|)\frac{d^2 u}{ d\tau^2}(\tau){\rm
d}\tau
\end{equation}
avec le Laplacien unidimensionnel $\frac{d^2 u}{ dx^2}$ et le noyau

\begin{equation}
\label{kernel} \ds
g(|x|)=\frac{h^{\delta}}{\delta(\delta-1)\epsilon}\,
|x|^{1-\delta} , \hspace{1cm} \delta \neq 1
\end{equation}
et avec $g(|x|)=-\frac{h}{\epsilon}\ln{|x|}$ dans le cas $\delta=1$.
De plus, il conviendrait d'exprimer (\ref{reforme}) en termes d'int\'egrales
fractionnelles soit

\begin{equation}
\label{fractionalaplaican} \ds
\Delta_{(\delta=2-D,\epsilon,h)}u(x)\approx \frac{h^{2-D}}{\epsilon}
\frac{\Gamma(D)}{(D-1)(D-2)}\left({\cal
D}_{-\infty,x}^{-D}+(-1)^D{\cal
D}_{\infty,x}^{-D}\right)\Delta_1u(x)
\end{equation}
o\`u $\Delta_1 u(x)=\frac{d^2}{dx^2}u(x)$ d\'enomme encore le Laplacien
unidimensionnel et on a introduit $D=2-\delta>0$ qui est positif dans l'intervalle
admissible $0<\delta<2$. Nous verrons par la suite que $D$ peut \^etre d\'efinie
dans l'intervalle $0<\delta<1$ comme la dimension fractale estim\'ee \cite{Hardy} de la courbe de la relation de dispersion correspondant \`a notre Laplacien.

Dans (\ref{fractionalaplaican}) on avait introduit l'int\'egrale fractionnelle de
{\it Riemann-Liouville} ${\cal
D}_{a,x}^{-D}$ laquelle est d\'efinie par (voir par exemple \cite{miller,kilbas})

\begin{equation}
\label{fractional} {\cal
D}_{a,x}^{-D}v(x)=\frac{1}{\Gamma(D)}\int_a^x(x-\tau)^{D-1}v(\tau){\rm
d}\tau
\end{equation}
o\`u ${\Gamma(D)}$ indique la fonction $\Gamma$ qui repr\'esente
une g\'en\'eralisation de la fonction factorielle aux nombres non-entiers positifs $D>0$.
La fonction $\Gamma$ est d\'efinie par

\begin{equation}
\label{gamma} \Gamma(D)=\int_0^{\infty}\tau^{D-1} e^{-\tau} {\rm d}
\tau \,,\hspace{1cm} D > 0
\end{equation}

Pour des entiers positifs $D>0$, la fonction de $\Gamma$ reproduit
la fonction factorielle  $\Gamma(D)=(D-1)!$ avec $D=1,2,..\infty$.
\newline\newline


\section{Le mod\`ele physique}
\label{thomas2}

On consid\`ere une cha\^ine lin\'eaire d'une distribution quasi-continue
de particules. Chaque point spatial $x$ correspond \`a un ``point mat\'eriel"
c'est \`a dire une particule. La densit\'e de masse suppos\'ee homog\`ene
est \'egale \`a un \`a chaque point de masse. Chaque particule est associ\'ee \`a
un seul degr\'e de libert\'e repr\'esent\'e par le champ $u(x,t)$ o\`u $x$
est la coordonn\'ee Lagragienne et $t$ le temps.
Chaque particule localis\'ee \`a $x$ est non-localement connect\'ee
par des ressorts d'intensit\'e $\xi^s$ ($\xi=N^{-\delta}, N\in \R >1, \delta >0$)
aux autres particules qui se trouvent aux positions $x\pm N^s h$
($h>0$), avec $s=0, \pm 1,\pm 2,..\pm\infty $. Le Hamiltonien d'une telle cha\^ine s'\'ecrit

\begin{equation}
\label{Hamiltonian}
H=
\frac{1}{2}\int_{-\infty}^{\infty}\left(\dot{u}^2(x,t)+{\cal V}(x,t,h)\right){\rm d}x
\end{equation}
Dans l'esprit de (\ref{self-sim-func}) l'\'energie \'elastique
${\cal V}(x,t,h)$ est construite de mani\`ere auto-similaire, soit
\footnote{Le facteur suppl\'ementaire ${1}/{2}$ \'evite un double comptage.}

\begin{equation}
\label{homchain} {\cal V}(x,t,h)=\frac{1}{2}{\hat
T}_N(h)\left[(u(x,t)-u(x+h,t))^2 +(u(x,t)-u(x-h,t))^2\right]
\end{equation}
o\`u ${\hat T}_N(h)$ est l'op\'erateur auto-similaire de (\ref{selfsimop}).
On arrive donc avec $\xi=\Lambda^{-1}=N^{-\delta}$ \`a

\begin{equation}
\label{elastic}
{\cal V}(x,t,h)=\frac{1}{2}\sum_{s=-\infty}^{\infty}\xi^s\left[(u(x,t)-u(x+hN^s,t))^2
+(u(x,t)-u(x-hN^s,t))^2\right]
\end{equation}
L'\'energie \'elastique remplit donc la condition d'auto-similarit\'e
par rapport \`a $h$, soit

\begin{equation}
\label{selfsimelasr}
{\hat A}_N(h){\cal V}(x,t,h)={\cal V}(x,t,Nh)= \xi^{-1}{\cal V}(x,t,h)
\end{equation}

L'exigence que l'\'energie \'elastique soit finie entra\^ine
la n\'ecessit\'e de convergence de la s\'erie (\ref{elastic}).
Pour cela on trouve pour l'exposant $\delta$ la contrainte \cite{michel-et-al}
\begin{equation}
\label{interval2}
0 < \delta < 2
\end{equation}

Afin d'\'etablir (\ref{interval2}) on a utilis\'e \'egalement le fait que $u(x,t)$ est Fourier-transformable. L'in\'equation (\ref{interval2}) d\'etermine l'intervalle o\`u l'\'energie
\'elastique (\ref{elastic}) converge.
L'\'equation de mouvement s'obtient par
\begin{equation}
\label{eqmo} \frac{\partial^2 u}{\partial t^2}=-\frac{\delta
H}{\delta u}
\end{equation}
(o\`u $\delta . / \delta u$ indique la d\'eriv\'ee fonctionnelle) soit

\begin{eqnarray}
\frac{\partial^2 u}{\partial t^2} &=&
- \sum_{s=-\infty}^{\infty} \xi^s \left( 2u(x,t)-u(x+hN^s,t)-u(x-hN^s,t) \right)
\label{eqmo2a}
\\
\frac{\partial^2 u}{\partial t^2} &=& \Delta_{(\delta,N,h)}u(x,t)
\label{eqmo2}
\end{eqnarray}
avec le Laplacien auto-similaire $\Delta_{(\delta,N,h)}$ de l'\'equation (\ref{lapla}).
Comme (\ref{elastic}), l'\'equation du mouvement (\ref{eqmo2a}) existe si $\delta$ est dans l'intervalle
(\ref{interval2}).
On peut r\'e\'ecrire (\ref{eqmo2}) sous la forme compacte d'une \'equation d'ondes

\begin{equation}
\label{comp} \Box_{\small(\delta,N,h)}u(x,t)=0
\end{equation}
o\`u $\Box_{\small(\delta,N,h)}$ est {\it l'auto-similaire analogue
de l'op\'erateur d'Alembertien d'ondes} ayant la forme

\begin{equation}
\label{comp2}
\Box_{\small(\delta,N,h)}=\Delta_{(\delta,N,h)}-\frac{\partial^2}{\partial
t^2}
\end{equation}
Le d'Alembertien (\ref{comp2}) avec le Laplacien (\ref{lapla})
d\'ecrit la propagation d'ondes dans la cha\^ine auto-similaire
dont le Hamiltonien est donn\'e par (\ref{Hamiltonian}).
Il semble que cette approche puisse constituer le point de d\'epart
pour une th\'eorie g\'en\'erale de la propagation d'ondes dans des milieux auto-similaires.

Le but suivant est de d\'eterminer la relation de dispersion, laquelle se constitue
par les valeurs propres (n\'egatives) du Laplacien semi-n\'egatif d\'efini (\ref{lapla}).
On utilise le fait que le champ $u(x,t)$ est Fourier-transformable et que la fonction exponentielle
$e^{ikx}$ est fonction propre du Laplacien (\ref{lapla}).
En \'ecrivant

\begin{equation}
\label{ufour} u(x,t)=\frac{1}{2\pi}\int_{-\infty}^{\infty}{\tilde
u}(k,t)e^{ikx}{\rm d}k
\end{equation}
on se rend compte que les amplitudes de Fourier ${\tilde
u}(k,t)$ satisfont

\begin{equation}
\label{eqmofou} 
\frac{\partial^2 \tilde{u}}{\partial t^2}(k,t)=-{\bar
\omega}^2(k)\,{\tilde u}(k,t)
\end{equation}

On obtient donc
\begin{equation}
\label{mandelbrot}{
\omega^2(kh)= 4\sum_{s=-\infty}^{\infty}N^{-\delta
s}\sin^2(\frac{khN^s}{2})}
\end{equation}

La s\'erie (\ref{mandelbrot}) d\'ecrit une fonction de type {\it Weierstrass-Mandelbrot} qui est continue et qui est pour $0<\delta \leq 1$ non-diff\'erentiable \cite{1,Hardy}. La fonction de Weierstrass-Mandelbrot satisfait toujours la condition d'auto-similarit\'e (\ref{mandelbrot}), soit
\begin{equation}
\label{selfgrap} \omega^2(Nkh)=N^{\delta}\,\omega^2(kh)
\end{equation}
sur la totalit\'e de son intervalle de convergence (\ref{interval2}).

On souligne le fait que seuls les exposants $\delta$
dans l'intervalle (\ref{interval2}) sont admissible pour l'Hamiltonien
(\ref{Hamiltonian}) pour d\'efinir un probl\`eme bien pos\'e.
Il fut d\'emontr\'e par Hardi \cite{Hardy} que pour $\xi N>1$ avec
$\xi=N^{-\delta} <1$, c'est \`a dire
\begin{equation}
\label{deltaconv2} 0 < \delta < 1
\end{equation}
la fonction de Weierstrass-Mandelbrot (\ref{mandelbrot})
est aussi une courbe fractale ayant une dimension fractale (dimension de
Hausdorff) $D=2-\delta > 1$. Les figures 2-4 montrent des courbes de dispersion
pour des valeurs d\'ecroissantes de $\delta$ dans l'intervalle $0< \delta < 1$.
Il est bien visible que la d\'ecroissance de $\delta$ produit une croissance
de la dimension fractale $D$. La hausse de la dimension fractale de la figure 2
\`a la figure 4 est bien visible par leur comportement de plus en plus turbulent.
Dans la figure 4, la dimension fractale est $D=1.9$ d\'ej\`a pr\`es de la dimension 2
du plan. Par contre, la figure 1 correspond avec $\delta=1.2$ \`a un cas non-fractal.
Afin d'\'evaluer (\ref{mandelbrot}) approximativement il convient de remplacer la s\'erie
par une int\'egrale en utilisant la m\^eme approximation qu'au paragraphe
\ref{contapprox} ($\epsilon\approx \ln N$). Dans le cadre de cette approximation
on trouve pour $|k|h$ suffisamment ``petit"  ($h>0$), c'est \`a dire dans le cas limite d'ondes
longues

\begin{equation}
\label{contapp} \omega^2(kh)\approx
\frac{(h|k|)^{\delta}}{\epsilon}C
\end{equation}
o\`u $(|k|h)^{\delta}$ doit \^etre de l'ordre de grandeur de $\epsilon$ ou plus petit.
La relation de dispersion est donc caract\'eris\'ee dans le cas limite d'ondes longues, par une loi de puissance
de la forme de ${\bar \omega}(k)\approx Const\, |k|^{\delta/2}$.
La constante $C$ qui est introduite dans (\ref{contapp})
est donn\'ee par l'int\'egrale

\begin{equation}
\label{CCabrev}
C=2\int_0^{\infty}\frac{(1-\cos{\tau})}{\tau^{1+\delta}}{\rm d}\tau
\end{equation}
laquelle existe uniquement si $\delta$ est dans l'intervalle (\ref{interval2}).
Dans ce cas limite, on obtient la densit\'e d'oscillateurs par \cite{michel-et-al}

\begin{equation}
\label{oscden} \rho(\omega)= 2\frac{1}{2\pi}\frac{d |k|}{d\omega}
\end{equation}
qui est normalis\'e de sorte que $\rho(\omega){\rm d}\omega$ compte le nombre d'oscillateurs, c'est \`a dire le nombre d'ondes qui se propagagent avec une certaine fr\'equence, laquelle se trouve dans
l'intervalle  $[\omega,\omega+{\rm d}\omega]$. On obtient \'egalement pour la densit\'e d'oscillateurs
une loi de puissance sous la forme de

\begin{equation}
\label{oscden2} \rho(\omega)= \ds \frac{2}{\pi\delta
h}\left(\frac{\epsilon}{C}\right)^{\frac{1}{\delta}}\omega^{\frac{2}{\delta}-1}
\end{equation}
avec $\delta$  toujours dans l'intervalle (\ref{interval2}). On observe donc que
la puissance $\frac{2}{\delta}-1$ est restreinte \`a l'intervalle $0< \frac{2}{\delta}-1 < \infty$
pour $0<\delta<2$.
En particulier, on trouve l'annulation de la densit\'e d'oscillateurs
pour une fr\'equence nulle $\rho(\omega\rightarrow 0)=0$.

\section{Conclusions}

Nous avons construit des fonctions et op\'erateurs de mani\`ere auto-similaire
en utilisant une certaine cat\'egorie de fonctions traditionnelles admissibles.
Avec cette approche il est possible de construire une sorte de
Laplacien auto-similaire et par cons\'equent de l'op\'erateur d'Alembertien d'ondes.
A l'aide de cette machinerie on d\'eduit une equation de mouvement qui d\'ecrit la propagation d'ondes
dans une cha\^ine auto-similaire. Nous croyons que ce-mod\`ele est l'un des plus simple
pour prendre en compte l'aspect d'auto-similarit\'e d'un milieux.
Nous d\'emontrons \'egalement la correspondance entre notre approche et le calcul fractionnel
qui fut employ\'e plus t\^ot \cite{6} afin d'aborder des probl\`emes physiques dans des milieux fractals.

Nous esp\`erons que notre approche pourra servir de point de d\'epart pour d\'evelopper
une th\'eorie de la propagation d'ondes dans des milieux auto-similaires et fractals
dans des contextes pluridisciplinaires.

\section{Acknowledgements}
Les auteurs sont reconnaissants \`a J.-M. Conoir et \`a D. Queiros-Conde pour des discussions
profondes.

\begin{figure}[t]
\psfig{figure=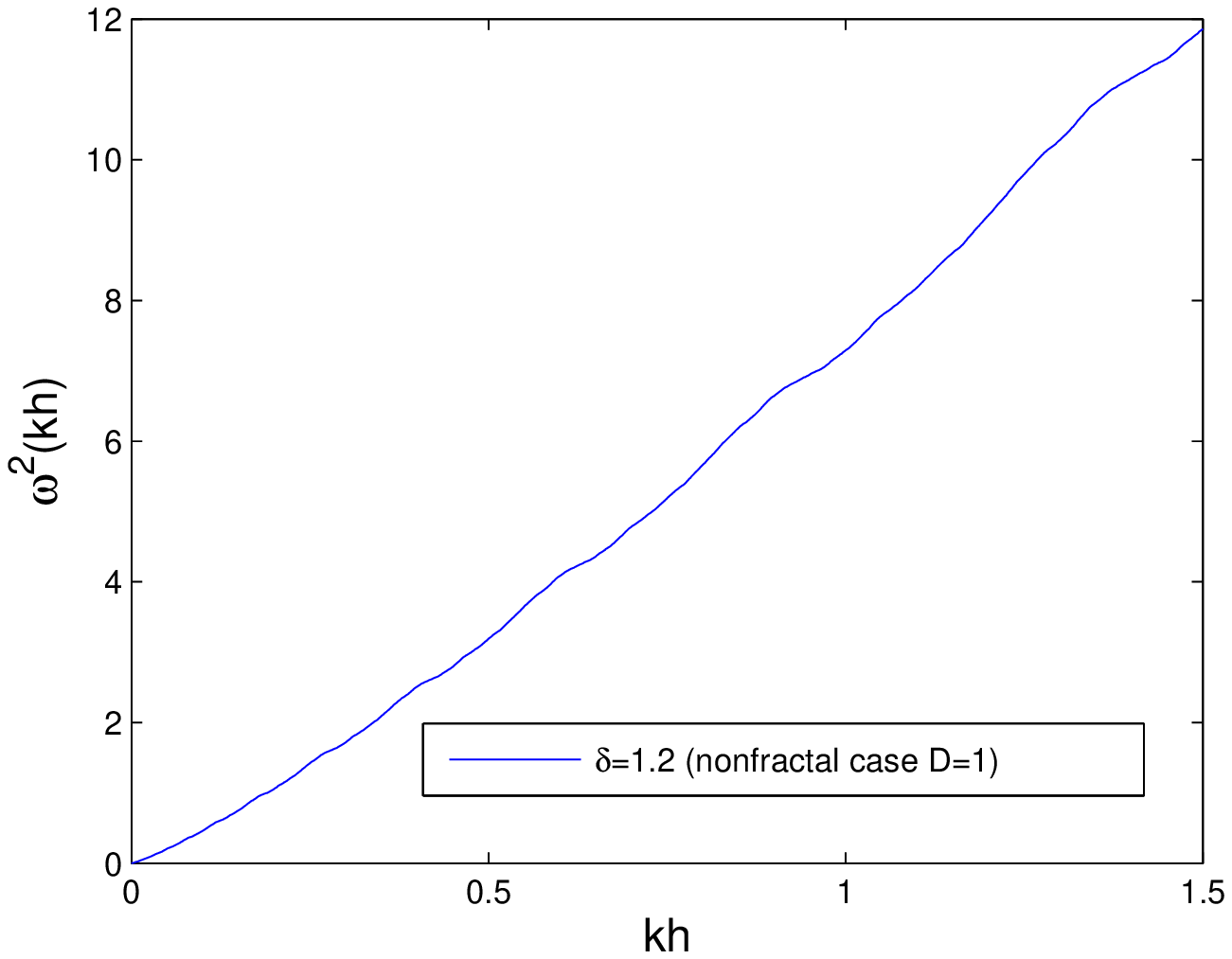,height=4in,width=4in}
\caption{Relation de dispersion $\omega^2(kh)$ en unit\'es arbitraires avec $N=1.5$ et $\delta=1.2$ }
\psfig{figure=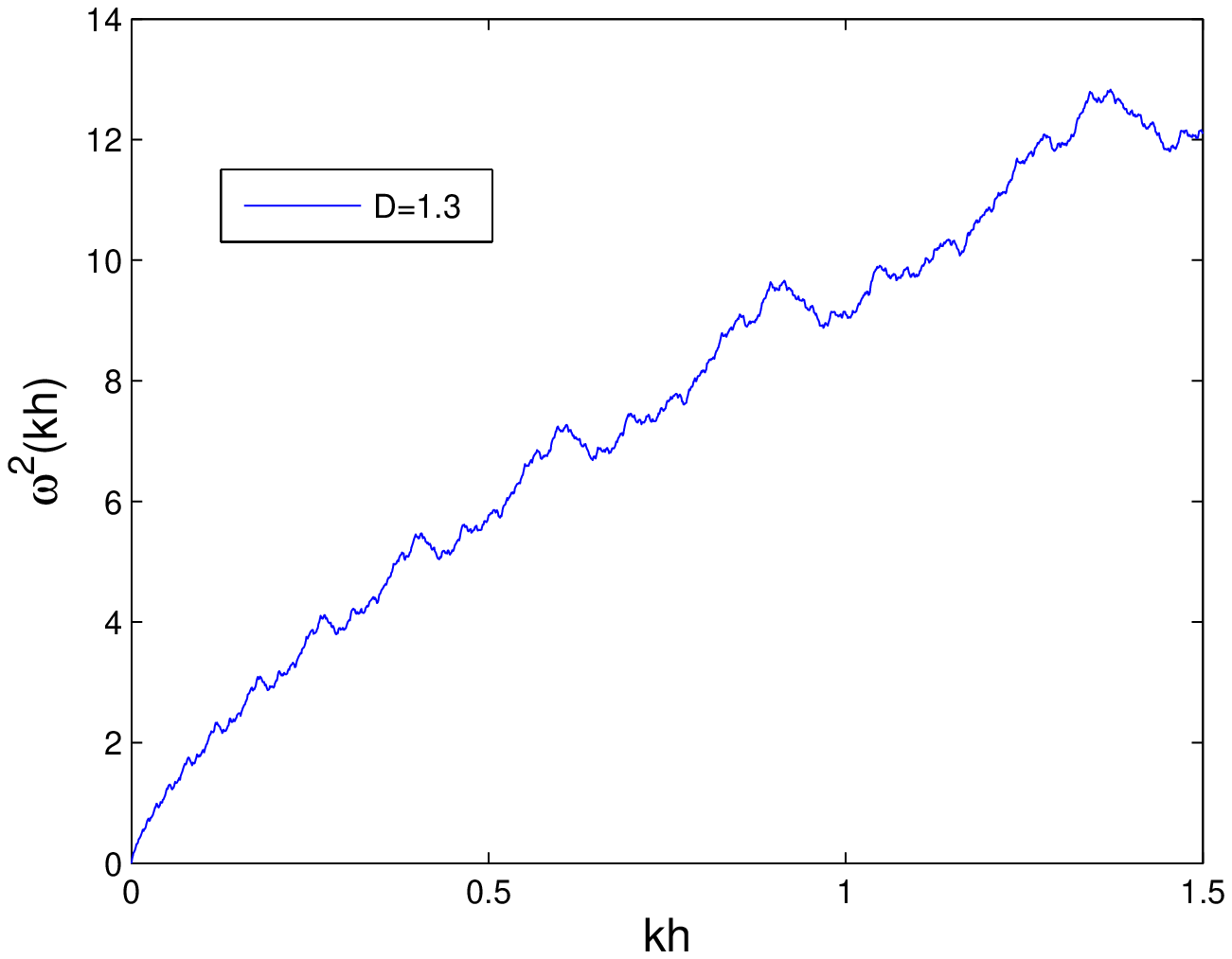,height=4in,width=4in}
\caption{Relation de dispersion $\omega^2(kh)$ en unit\'es arbitraires avec $N=1.5$ et $\delta=0.7$}
\end{figure}
\begin{figure}[t]
 \psfig{figure=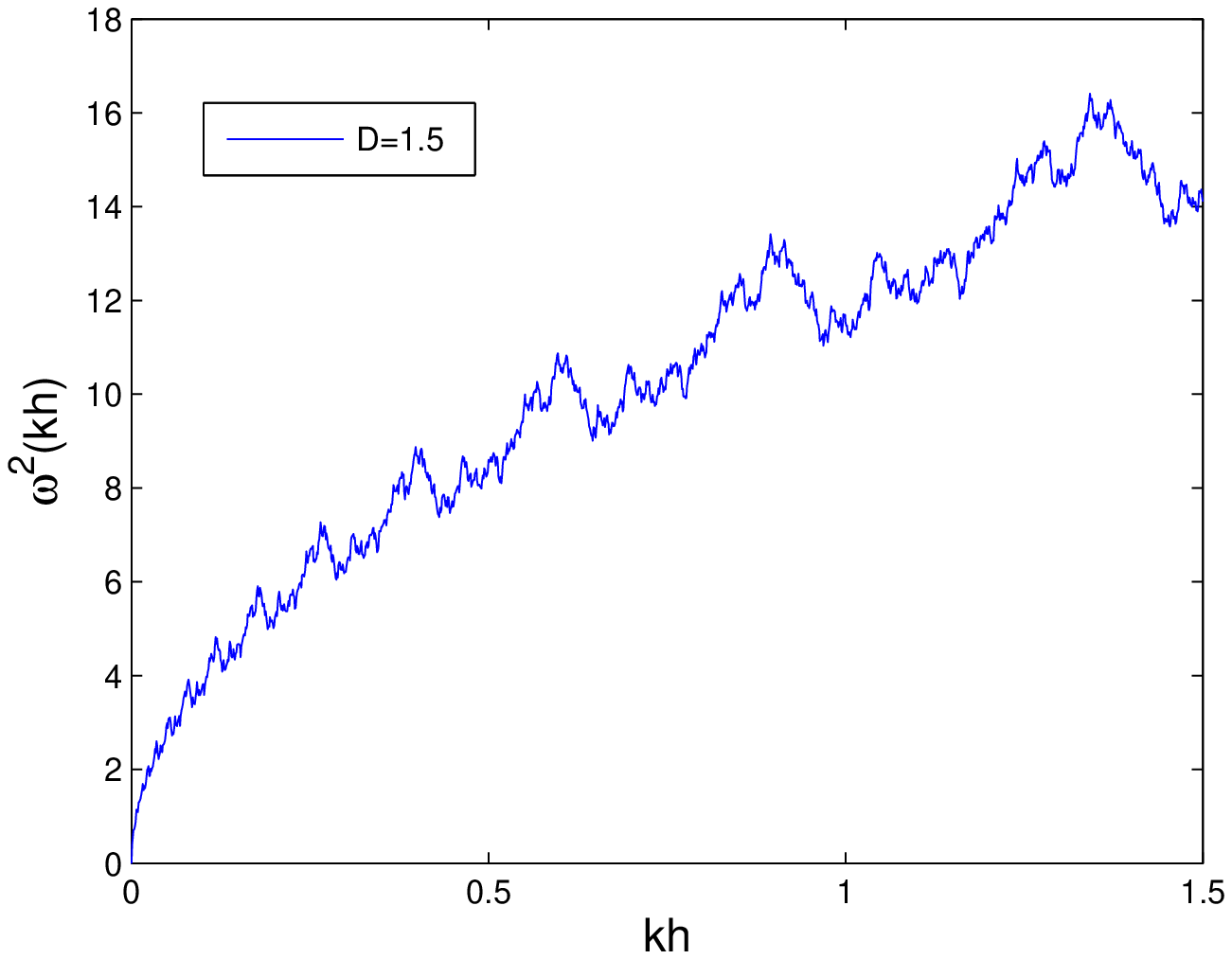,height=4in,width=4in}
 \caption{Relation de dispersion $\omega^2(kh)$ en unit\'es arbitraires avec $N=1.5$ et $\delta=0.5$}
\psfig{figure=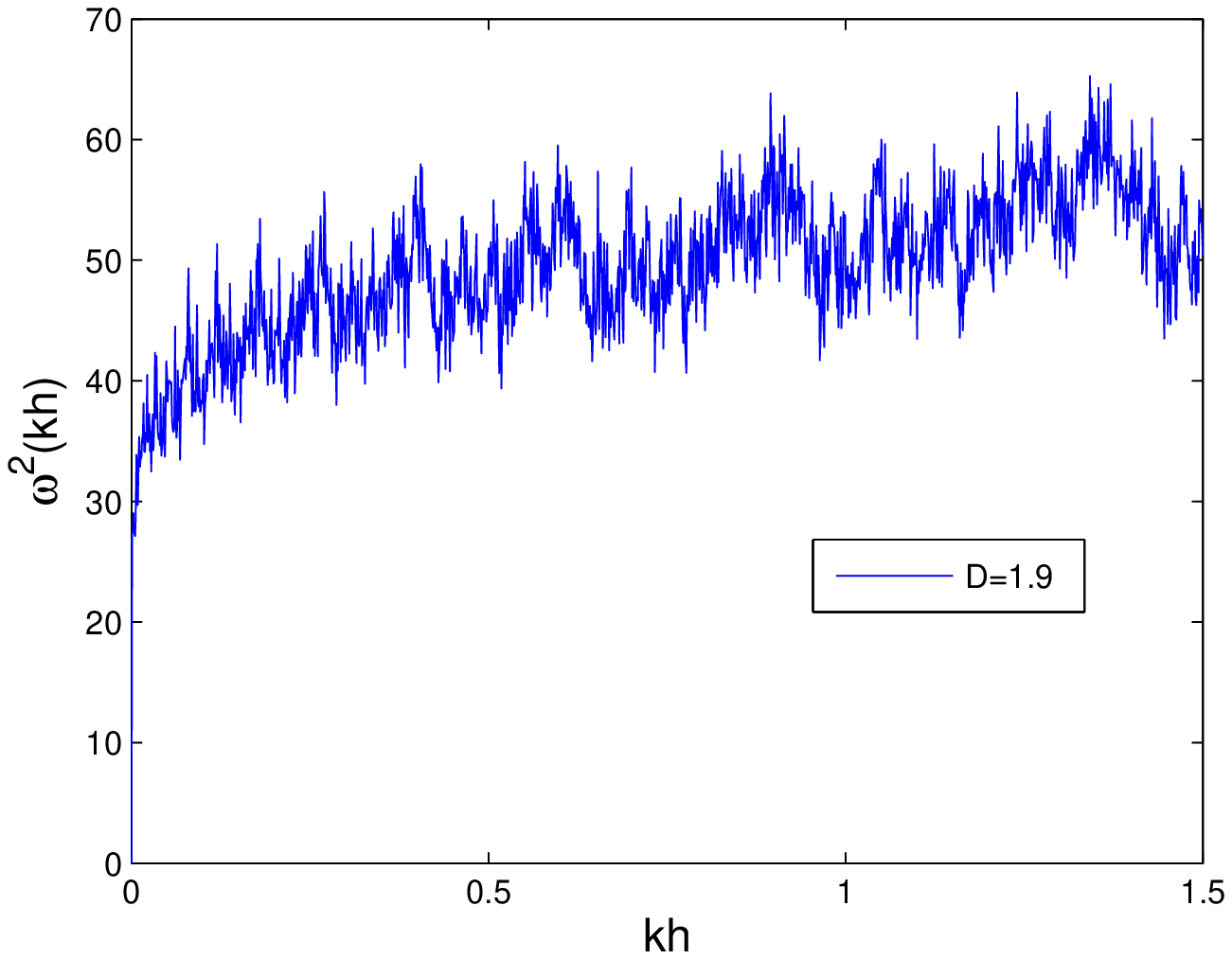,height=4in,width=4in}
\caption{Relation de dispersion $\omega^2(kh)$ en unit\'es arbitraires avec $N=1.5$ et $\delta=0.1$}
\end{figure}

\bibliographystyle{spmpsci}

\end{document}